\documentclass[%
 reprint,
 amsmath,amssymb,
 aps,
]{revtex4-2}

\usepackage{graphicx}% Include figure files
\usepackage{dcolumn}% Align table columns on decimal point
\usepackage{bm}% bold math
\usepackage{dsfont}
\usepackage{amsmath}
\usepackage{amssymb}
\usepackage{float}
\usepackage{notoccite}
\usepackage[svgnames]{xcolor}
\usepackage{physics}
\usepackage{mathrsfs}
\usepackage{accents}
\usepackage{empheq}
\usepackage{amsthm}
\usepackage{appendix}
\begin{document}

\preprint{APS/123-QED}

\title{Time of occurrence observables:\\expanding to other symmetries}% Force line breaks with \\

\author{V. Cavalheri Pereira}
 \email{vitor.cavalheri.pereira@usp.br}
\author{J. C. A. Barata}%
\affiliation{%
 Instituto de Física da Universidade de São Paulo, Rua do Matão 1371, São Paulo, Brazil\\
}%


%\affiliation{
%\affiliation{
 %Third institution, the second for Charlie Author
%\affiliation{%


\date{\today}% It is always \today, today,

\begin{abstract}
Recent works have proposed the use of the formalism of Positive Operator Valued Measures to describe time measurements in quantum mechanics. This work aims to expand on the work done by other authors, by generalizing the previously proposed construction method of such measures to include causal Poincaré transformations, in order to construct measures which are covariant with respect to such transformations.
%\b order toegin{description}
%\end{description}
\end{abstract}

\maketitle

%\tableofcontents

\section{\label{sec:level1}Introduction}

The question of how to interpret time in quantum mechanics has been a topic of discussion since the subject's inception \cite{HeisenbergUncertainty}. But it has been shown \cite{Pauli} that it is impossible to interpret time as an usual physical observable, with the foundations laid out at the time. A new set of axioms was suggested in \cite{Ludwig}, in which time could then be understood as a "generalized observable". This association was made by maintaining observables as associated to Projective Valued Measures (PVMs), and associating Positive Operator Valued Measures (POVMs) to generalized observables. This line of reasoning was followed in several works thereafter, by using it to construct time of arrival observables in many different situations \cite{Kijowski1974}, \cite{Kijowski1999}, \cite{Werner} to cite some works in this line of thought.\\
The route taken in \cite{Brunetti2002.1} was different from the previous works, in the sense that the authors laid out a general method of construction of POVMs, them being associated to specific observables being measured in a system, such that the first moment of this POVM could then be understood as a time of occurrence operator to this measured observable. Such POVM would be covariant with respect to time translations and an example was studied where a general formula for the time delay for a scattering process was calculated through much easier means than what was previously done. Such general method of construction of POVMs and time of occurrence operators could then be extended to other symmetries, in order to be used in a wider array of problems and contexts. This was the goal of this work, in which the method proposed in \cite{Brunetti2002.1} was extended in order that the constructed measures were covariant with respect to causal Poincaré transformations.\\
In section II, \cite{Brunetti2002.1} will be revisited in order to make the path of the generalization more clear. In section III, the method to constructing Poincaré covariant measures will be exposed. In section IV some conclusions will be drawn, and possible paths forward will be indicated.

\section{Mathematical Preliminaries}

As suggested in \cite{Brunetti2002.1}, the base for the 
proposed method is to analyse $\alpha_t(A)$, where $A$ is a positive and bounded observable and $\alpha_t$ is the action of the group of time translations. These properties forced upon the observable $A$ are necessary, due to the fact that the process of constructing these measures can be understood as a normalization done to the level of operators, with the resulting measure being a probabilistic one. By requiring the observable to be positive, the quantity $\alpha_t(A)$ becomes a non-negative function, and a measure constructed using it as a basis will thus not be a signed one. By requiring the observable to be bounded, the quantity $\alpha_t(A)$ becomes normalizable, and thus the measure constructed will be a probabilistic one. In the case of \cite{Brunetti2002.1}, the authors chose to focus on constructing a covariant measure with respect to time translations, in order to arrive at a time of occurrence operator that would behave correctly under such transformations. In the spirit of applying this method to a greater class of physical systems, with a particular interest in using this possibly in the context of localizability operators defined in \cite{Newton}, this work aims to follow the steps of \cite{Brunetti2002.1} in order to build POVMs covariant with respect to causal Poincaré transformations.\\
Starting now a review of \cite{Brunetti2002.1}, one begins by analyzing the following quantity: $\alpha_x(A)$, where $\alpha_x$ is the action of the translation group in $\mathbb{R}^n$, $(\mathbb{R}^n, +)$, the first coordinate being understood as a time coordinate. Since $A$ is positive and bounded, $\alpha_x(A)$ will thus be a non-negative, finite quantity, and from the integrals:

$$\int_I \langle \Psi, \alpha_x(A)\Phi \rangle dx$$
\\
with $I$ being Borel subsets of $\mathbb{R}^n$, we may define the operators:

\begin{equation}
    B(I) := \int_I \alpha_x(A) dx \; .
\end{equation}
\\
Here, $\Phi, \Psi$ are elements of Hilbert spaces where the action of the translation group is well defined.\\
Consider now the net of sets with the partial ordering $\{I_{\lambda}\}_{\lambda \in \Lambda}$, $I_{\lambda} \subset I_{\lambda'}$ if $\lambda \prec \lambda'$. It must be noted that such a net must also include all the translated sets of the form $I_{\lambda} + y$ for every $\lambda \in \Lambda$ and $y \in \mathbb{R}^n$. This net induces the same partial ordering on the $B(I_{\lambda})$, considering that $B(I)$ is a positive operator for every set $I$, and thus, if $\lambda \prec \lambda'$, it holds:

\begin{eqnarray}
    B(I_{\lambda'}) = \int_{I_{\lambda}} \alpha_x(A)dx + \int_{I_{\lambda'}\setminus I_{\lambda}} \alpha_x(A) dx = \nonumber\\
    B(I_{\lambda}) + B(I_{\lambda'}\setminus I_{\lambda}) \;. \nonumber
\end{eqnarray}
\\
That is, since $B(I_{\lambda'}\setminus I_{\lambda}) \geq 0$, then $B(I_{\lambda'}) \geq B(I_{\lambda})$. From this net of operators one can induce the following net of inverse of resolvent operators: $\{[\mathds{1} + B(I_{\lambda})]^{-1}\}_{\lambda \in \Lambda}$. A simple result as seen in \cite{BrattelieRobinson} (c.f. Proposition 2.2.13, item d) shows that these inverses are all well defined, positive and besides, that the ordering on this net is the inverse from the previous ones in the sense that, if $\lambda \prec \lambda'$, then $[\mathds{1} + B(I_{\lambda'})]^{-1} \leq [\mathds{1} + B(I_{\lambda})]^{-1}$.
\\
Now, a net of positive operators which is decreasing, will converge strongly to a positive operator which will be designated $C$. To see that, one may start from a given normalized element $\Psi$ of the Hilbert space $\mathscr{H}$ and due to the ordering on this net, it holds for this fixed $\Psi$:

$$|\langle \Psi, [\mathds{1} + B(I_{\lambda})]^{-1} \Psi \rangle - \langle \Psi, [\mathds{1} + B(I_{\lambda'})]^{-1} \Psi \rangle|^2 < \epsilon^2_{\Psi}$$
\\
for suitably chosen $\lambda, \lambda' \in \Lambda$. Defining $D_{\lambda, \lambda'} := [\mathds{1} + B(I_{\lambda, \lambda'})]^{-1}$, the previous equation is written as:

$$\langle \Psi, (D_{\lambda} - D_{\lambda'})\Psi \rangle < \epsilon_{\Psi}^2\;.$$
\\
Now focusing on the next expression, we obtain:

$$\norm{(D_{\lambda} - D_{\lambda'})\Psi}^2 = \langle \Psi, (D_{\lambda} - D_{\lambda'})^2\Psi \rangle \;,$$
\\
this expression resulting from the fact that both $D_{\lambda}$ and $D_{\lambda'}$ are self-adjoint, their subtraction being self-adjoint as well.\\
With the appropriate choice of $\lambda$ and $\lambda'$ such that $D_{\lambda'} \leq D_\lambda$, along with the fact that $D_{\lambda} \leq \mathds{1}$ for every $\lambda$, then it holds that $\norm{D_{\lambda} - D_{\lambda'}} < 1$, and once again by \cite{BrattelieRobinson} (c.f. Proposition 2.2.13 - item b) it can be concluded that:

$$1 \cdot (D_{\lambda} - D_{\lambda'}) \geq \norm{D_{\lambda} - D_{\lambda'}}(D_{\lambda - D_{\lambda'}}) \geq (D_{\lambda} - D_{\lambda'})^2 \geq 0$$
\\
This means then that:

\begin{eqnarray}
  \norm{(D_{\lambda} - D_{\lambda'})\Psi}^2 = \langle \Psi, (D_{\lambda} - D_{\lambda'})^2\Psi \rangle \nonumber\\
  \leq \langle \Psi, (D_{\lambda} - D_{\lambda'})\Psi \rangle < \epsilon_{\Psi}^2\;. \nonumber
\end{eqnarray}
\\
This is exactly the strong convergence of the net of operators $\{[\mathds{1} + B(I_{\lambda})]^{-1}\}_{\lambda \in \Lambda}$.
\\
There is also the question of the uniqueness of such a limit operator, since the set of operators being considered forms a net instead of a sequence. We may see that the positiveness of the operators in the net leads directly to the uniqueness of the limit operator. Consider there exists certain choices of indices in $\Lambda$, one converging to the operator $C$ and one converging to the operator $C'$. Take from these choices of indices, two particular indices $\lambda'$ and $\lambda''$, such that it holds

\begin{equation*}
  \left\{\begin{array}{@{}l@{}}
    \norm{(C - [\mathds{1} + B(I_{\lambda'})]^{-1})\Psi} < \frac{\epsilon_{\Psi}}{4} \; ,\\
    \norm{(C' - [\mathds{1} + B(I_{\lambda''})]^{-1})\Psi} < \frac{\epsilon_{\Psi}}{4} \; .
  \end{array}\right.\,
\end{equation*}
\\
Since $\Lambda$ is a directed set, there is an index $\lambda''' \in \Lambda$, such that $\lambda', \lambda'' \prec \lambda'''$. In the level of the net of operators, this means that $B(I_{\lambda'''}) \geq B(I_{\lambda'}), B(I_{\lambda''})$, which in particular implies
\begin{equation*}
    [\mathds{1} + B(I_{\lambda'''})]^{-1} \leq [\mathds{1} +
    B(I_{\lambda'})]^{-1}, [\mathds{1} + B(I_{\lambda''})]^{-1} \;.
\end{equation*}
\\
It holds true that 

\begin{equation*}
  \left\{\begin{array}{@{}l@{}}
    C \leq [\mathds{1} + B(I_{\lambda'''})]^{-1} \leq [\mathds{1} + B(I_{\lambda'})]^{-1} \; ,\\
    \norm{(C - [\mathds{1} + B(I_{\lambda'})]^{-1})\Psi} < \frac{\epsilon_{\Psi}}{4} \;.
  \end{array}\right.\,
\end{equation*}
\\
so, it follows in particular that 
\begin{equation*}
    \norm{([\mathds{1} + B(I_{\lambda'})]^{-1} - [\mathds{1} + B(I_{\lambda'''})]^{-1})\Psi} < \frac{\epsilon_{\Psi}}{4} \;.  
\end{equation*}
\\
The same holds for $\norm{([\mathds{1} + B(I_{\lambda''})]^{-1} - [\mathds{1} + B(I_{\lambda'''})]^{-1})\Psi}$.
\\
Denoting once again $D_{\lambda} := [\mathds{1} + B(I_{\lambda})]^{-1}$, we can see then:

\begin{eqnarray}
    \norm{(C - C')\Psi} \leq \norm{(C - D_{\lambda'})\Psi} + \norm{(D_{\lambda''} - C')\Psi} + \nonumber \\
    \norm{(D_{\lambda'} - D_{\lambda'''})\Psi} + \norm{(D_{\lambda'''} - D_{\lambda''})\Psi} \;.\nonumber
\end{eqnarray}
\\
That is, one has

\begin{equation}
    \norm{(C - C')\Psi} < \epsilon_{\Psi} \;.
\end{equation}
\\
This means that the limit operator is unique.
\\
This limit operator $C$ can be then used to properly define the main object of interest. As said previously, the matter of constructing these POVMs is a matter of knowing whether or not the constructed measures are normalizable. In other words, the issue is determining if:

\begin{equation}
    B := \int_{\mathbb{R}^n} \alpha_x(A)dx
\end{equation}
\\
is well defined. Considering the net for which the operator $C$ is the limit of, it is rather natural to the define:

\begin{equation}
    B:= C^{-1} - \mathds{1}
\end{equation}
\\
which leads to the question of where is $C^{-1}$ well defined. Two spaces will be considered for this, which will be denoted $\mathscr{H}_0$ and $\mathscr{H}_{\infty}$. These spaces are respectively the space formed by the joint kernel of the operators $B(I)$, and the kernel of the operator $C$. They are denoted such since as noted by the authors in \cite{Brunetti2002.1}, they will be the spaces corresponding to states where no event takes place, or the events take infinitely long to happen, respectively. In the orthogonal complement to these spaces, $\mathscr{H}_{fin.} = (\mathscr{H}_0 + \mathscr{H}_{\infty})^{\perp}$, the inverse of the operator $C$ will be well defined, and thus the operator $B$ will also be so.
\\
The operator $B$ can be shown to be self-adjoint, positive, and most importantly, it dominates all operators $B(I)$, thus also dominating their translations. Through the use of the spectral theorem, the operators $B^{-\frac{1}{2}}$ are directly defined, and one can thus define:

\begin{equation}
    P(I) := B^{-\frac{1}{2}}B(I)B^{-\frac{1}{2}} \;.
\end{equation}
\\
This operator, as defined, is bounded from above by the identity, and it is easy to see that it also obeys finite additivity. Through the topological properties of second countable spaces, such as $\mathbb{R}^n$, along with the previous finite additivity, $\sigma$-additivity follows through at once. At this point, it is easy to see that this measure has the covariance property with respect to translations.
\\
To show that $P(I)$ is covariant is to show that $\alpha_y(P(I)) = P(I + y)$ for any $I \subset \mathcal{B}(\mathbb{R}^n)$ and any $y \in \mathbb{R}^n$. Starting from a translation on $P(I)$:
        
        \begin{equation*}
            \alpha_y(P(I)) = \alpha_y(B^{-\frac{1}{2}}B(I)B^{-\frac{1}{2}})\;.
        \end{equation*}
\\        
Due to Wigner's Theorem, it follows:
        
        \begin{eqnarray}
            \alpha_y(B^{-\frac{1}{2}}B(I)B^{-\frac{1}{2}}) = U_y^*(B^{-\frac{1}{2}}B(I)B^{-\frac{1}{2}})U_y = \nonumber \\ \alpha_y(B^{-\frac{1}{2}})\alpha_y(B(I))\alpha_y(B^{-\frac{1}{2}}) \;.\nonumber
        \end{eqnarray}
\\        
Studying $\alpha_y(B)$:
        
        \begin{eqnarray}
            \alpha_y(B) = \lim_{I \rightarrow \mathbb{R}^n} \int_I \alpha_y(\alpha_x(A)) dx = \lim_{I \rightarrow \mathbb{R}^n} \int_I \alpha_{y + x}(A) dx = \nonumber \\
            \lim_{I \rightarrow \mathbb{R}^n} \int_{I - y} \alpha_x(A) dx = \int_{\mathbb{R}^n} \alpha_x(A) dx = B \;.\nonumber
        \end{eqnarray}
\\        
Thus $B$ is invariant under translations $\alpha_y$. Since $B$ is invariant, this means that its spectral projectors will also be invariant to translations, that is, $B^{-\frac{1}{2}}$ is also invariant with respect to these transformations. Finally, for $B(I)$:
        
        \begin{eqnarray}
            \alpha_y(B(I)) = \alpha_y\left(\int_I \alpha_x(A) dx\right) = \int_I \alpha_y(\alpha_x(A)) dx = \nonumber \\ 
            \int_I \alpha_{y + x}(A) dx = \int_{I - y} \alpha_x(A) dx = B(I - y), \nonumber
        \end{eqnarray}
\\        
and thus:
        
        \begin{equation}
            \alpha_y(P(I)) = B^{-\frac{1}{2}}B(I - y)B^{-\frac{1}{2}} = P(I - y) \;,
        \end{equation}
\\        
that is, $P(I)$ is covariant with respect to translations on $\mathbb{R}^n$.
\\
With the measure being constructed, we may then analyze the first moment of the first coordinate:

\begin{equation}
    T_A = \int_{\mathbb{R}^n}tP(dx) \;.
\end{equation}
\\
This can be interpreted as the time of occurrence operator associated to the event measured by observable $A$. We have thus constructed a measure, and through it an operator, which tracks the moment when the event measured by observable $A$ happens, and that transforms accordingly with respect to translations. Inspired by the ideas introduced in \cite{Brunetti2002.1}, we discuss in the next section the idea of laying out this method for more symmetries, specifically for causal transformations.

\section{A Method for Poincaré Covariant Measures}

The task of extending the previous construction to more symmetries boils down to the question of understanding how the operators

\begin{equation}
    B(V) = \int_V \alpha_g(A) d\mu(g)
\end{equation}
\\
can be defined for a given action $\alpha_g$ of a group $G$ on the chosen observable $A$, on a neighborhood $V$ of the group.\\
The structure of neighborhoods and integrals imply the necessity of at least some sort of topological structure to the groups where the previous construction method can be extended. Thus, attention in this work will be restricted to topological groups. In particular, the topological aspects of such groups allow for the study of defining measures on them, through Haar integrals.\\
Inspired by Lebesgue measures in $\mathbb{R}^n$, the concept of invariant measures was established, with the necessity of separating between \textbf{left} and \textbf{right} invariant measures \cite{Barut}. This is so since, given a measure $\mu$ on a group $G$, and two elements $h, g \in G$, it is not necessarily true that $\mu(gh) = \mu(hg)$. A left invariant measure is such that given an element $g \in G$, and a neighborhood $V \subset G$, then it holds $\mu(gV) = \mu(V)$. For a right invariant measure the previous equation holds with the change of $g$ acting on $V$ from the right.\\
By Haar's theorem \cite{Haar}, \cite{Nachbin}, the existence of left invariant measures is guaranteed for every locally compact group. It is rather straightforward to prove that this implies the existence of right invariant measures induced from these left invariant ones. If a Haar measure is simultaneously left and right invariant, it is said to be simply an invariant measure.
To understand how this method of constructing POVMs may be extended to the case of causal transformations, it is better to consider the structure of the group which contains these transformations. It is an elementary fact \cite{Barut} that the Poincaré group is the semidirect product between the additive group and the Lorentz group. Considering that the focus of this work is with respect to causal transformations, we will restrict ourselves to working with the causal Poincaré transformations, that is, the ones that preserve the interval between points of the space. Due to a result that will be enunciated at the proper time, it is enough to separate the analysis between each of the component subgroups of the product. Since the study for the translation group was already done in \cite{Brunetti2002.1} and revisited in the previous section, we will now draw parallels between it and the proper Lorentz group case.
\\
A first point that should be attended to is the question of the Haar measure defined on the proper Lorentz group. As was mentioned, there can be left or right invariant Haar measures, with them being not necessarily equal. In this particular work, invariant measures would be of greater interest than only left or only right invariant measures, since such measures have their associated right (or left) hand moduli, respectively. Due to an elementary result, invariant measures are such that their left and right handed moduli are identically equal to 1, which is particularly useful when dealing with processes such as normalizations, which is our current case. Groups which obey the previous property are thus said to be \textbf{unimodular} groups. The following theorem \cite{Barut} characterizes which groups are unimodular:\\

\textbf{Theorem 1:} The following groups are unimodular:

\begin{itemize}
    \item Lie Groups for which the set of modular functions $\{\Delta_r(g), g \in G\}$ is compact;
    \item Semisimple Lie Groups;
    \item Connected Nilpotent Lie Groups.
\end{itemize}

What is meant by a semisimple Lie group is simply a Lie group which contains no proper invariant connected abelian Lie group. As for a nilpotent Lie group is a Lie group such that, for the chain of sets $K_n$, with the elements of these sets being finite products of elements of the form $xyx^{-1}y^{-1}$, where $x \in K_{n - 1}$ and $y \in G$, $K_0 = G$, it holds that for a finite number $m \in \mathbb{N}$, $K_m = \{e\}$.
\\
It is a straightforward exercise to show that the proper ortochronous Lorentz group is a semisimple Lie group, and thus, by the previous theorem, it is also is a unimodular group. At this point the measure of the proper ortochronous Lorentz group may be simply referred to as $\mu(\Lambda)$, without differentiation as to whether it is left or right invariant. The interesting part in this process is that after taking the proper care of using invariant measures, most of the differences to the translation case are now gone. To see that, let us then start with the following operator:

\begin{equation}
    B(V) = \int_V \alpha_{\Lambda}(A) d\mu(\Lambda) \;,
\end{equation}
\\
where $V$ is a set of finite measure in the restricted Lorentz group, and $\alpha_{\Lambda}$ is the action of the Lorentz group on a bounded, positive observable $A$. From this point, the process is completely similar to what was done in the translation case. The operators $B(V)$ will all be positive, and we may consider the net formed by them $\{B(V_{\gamma})\}_{\gamma \in \Gamma}$, where $\Gamma$ is a family of indices, and such that this net also contains all of the Lorentz transformed operators of the form $\alpha_{\Lambda}B(V)$. The net $\{[\mathds{1} + B(V_{\gamma})]^{-1}\}_{\gamma \in \Gamma}$ is once again induced, having a unique limit operator $C$ due to being a descending net of positive operators, as before.\\
With this limit operator, we may then restrict ourselves to working in the space where the inverse of such an operator is defined. This space will be denoted as in the translation case $\mathscr{H}_{fin.} = (\mathscr{H}_{0} + \mathscr{H}_{\infty})^{\perp}$. In this space, properties about the inverse operator can be demonstrated such as before. In particular, it is a self-adjoint operator. With this, we may once again define:
\begin{equation}
    P(V) := B^{-\frac{1}{2}}B(V)B^{-\frac{1}{2}}
\end{equation}
\\
and, as before, use the finite additivity of the operator $P$, along with the fact that, by virtue of the proper Lorentz group being a Lie group, it is in particular a second countable space, to extend the domain of $P$ to include sets of non finite measure in $L^{\uparrow}_+$. With this, the operator $P$ thus described is a proper measure.
\\
More important is the fact that this operator will be covariant with respect to proper Lorentz transformations:
\begin{eqnarray}
\alpha_{\Lambda}(B(V)) = \int_V \alpha_{\Lambda}(\alpha_{\Lambda'}(A)) d\mu(\Lambda) =\nonumber\\
\int_V \alpha_{\Lambda\Lambda'}(A) d\mu(\Lambda') = \int_{\Lambda^{-1}V} \alpha_{\Lambda'}(A) d\mu(\Lambda') = B(\Lambda^{-1}V) \;.\nonumber
\end{eqnarray}
\\
Thus:
        
\begin{equation}
    \alpha_{\Lambda}(P(V)) = B^{-\frac{1}{2}}B(\Lambda^{-1}V)B^{-\frac{1}{2}} = P(\Lambda^{-1}V)\;.
\end{equation}
\\        
With this measure $P$ being constructed, the task is now to use it and the translation covariant measure to construct the full causal Poincaré measure. The following result shows how to do so \cite{Nachbin}:
\\

\textbf{Proposition:} Let a group $G$ be the semidirect product of two locally compact groups $H$ and $N$, with an action $\alpha_n$ of $N$ on $H$. Let $d\mu(h)$ and $d\nu(n)$ be the left invariant Haar measures defined on $H$ and $N$, respectively. Then, the left invariant Haar measure on $G$, denoted $d\xi(g)$ is given by:

\begin{equation}
    \int f(x)d\xi(g) = \int\int \frac{f(h, n)}{\delta^H(\alpha_n)}d\mu(h)d\nu(n) \;,
\end{equation}
\\
where $f$ is an integrable function defined on $G$. It also holds:

\begin{equation}
    \Delta_r^{G}(u, v) = \frac{\Delta_r^H(u)\Delta_r^N(v)}{\delta^H(\alpha_v)} \;,
\end{equation}
\\
where the superscripts indicate which group the object is referring to, and $\delta$ being the modulus of the topological automorphism $\alpha_{v}$.
\\
It is also direct to show that the causal Poincaré group is a connected nilpotent group, and through Theorem 1 will also be unimodular. Using the previous proposition, we obtain directly that the measure for the causal Poincaré group is given by:

\begin{equation}
    d\nu(\{a, \Lambda\}) = dad\mu(\Lambda)
\end{equation}
\\
With the measure being established, the same process may now be applied. We consider operators of the form:

\begin{equation}
    B(M) := \int_M \alpha_{\{a, \Lambda\}}(A) dad\mu(\Lambda)
\end{equation}
\\
where $M$ is a set of finite measure of the causal Poincaré group and and $A$ is a positive, bounded observable. Writing $M = I \cross V$, with $I$ a Borel set in the translation group and $V$ a Borel set in the causal Lorentz group, then we may rewrite the previous equation as:

\begin{eqnarray}
    B(M) = \int_I \int_V U^*(a, \Lambda)AU(a, \Lambda) dad\mu(\Lambda) \nonumber \\
    = \int_V U^*(0, \Lambda) \left[\int_I U(a, \mathds{1})^*AU(a, \mathds{1}) da\right] U(0, \Lambda) d\mu(\Lambda) \;. \nonumber \\
\end{eqnarray}
\\
Since the representations $U(0, \Lambda)$ and $U(a, \mathds{1})$ are equivalent to representations on the proper ortochronous Lorentz group and the translation group, respectively, then we may apply the previous method twice, in order to obtain a measure which is covariant with respect to causal Poincaré transformations. The main result of this work then follows through directly, with the covariant measure being given by:
\begin{equation*}
    P(M) := B^{-\frac{1}{2}} \left[\int_M \alpha_{\{a, \Lambda\}}(A) d\nu(\{a, \Lambda\})\right] B^{-\frac{1}{2}} =
\end{equation*}

\begin{eqnarray}
    = B^{-\frac{1}{2}} \left[\int_V U(0, \Lambda)^* \left(\int_I U(a, \mathds{1})^*AU(a, \mathds{1}) da\right) \right. \nonumber \\
    \times U(0, \Lambda) d\mu(\Lambda)\bigg]B^{-\frac{1}{2}} \;,
\end{eqnarray}
\\
and the associated time operator given as the first moment of the first coordinate of the translation group:\\

\begin{equation}
    T_A = \int_{\mathbb{R}^4}\int_{L_+^{\uparrow}} t P(d\{a, \Lambda\}) \;.
\end{equation}
\\
At this point, we have reached the goal of this work, which was to construct a time operator which behaved covariantly with respect to causal Poincaré transformations. There is still though, a matter of technical character which would be interesting to be touched upon. As is known \cite{Barut}, some representations of the Lorentz group are not considered true representations. To avoid having to deal with the modulus 1 factor appearing in these representations when doing explicit calculations, it would be better to maintain the work only with true representations. A seminal work by Bargmann showed that this could be done by lifting the work from the group where there are projective representations to its universal covering group.
\\
In this case, the universal covering of the Lorentz group is given by the group of 2 by 2 complex matrices with determinant 1: $SL(2, \mathbb{C})$. An arbitrary element of this group will be given as:

$$S = \begin{bmatrix}
    \alpha & \beta\\
    \gamma & \delta
  \end{bmatrix}$$\\
\\
It is easy to show that $SL(2, \mathbb{C})$ is a unimodular group. Its Haar measure will be given as:

\begin{equation}
    d\mu(S) = \frac{1}{|\delta|^2}d\beta d\Bar{\beta} d\gamma d\Bar{\gamma} d\delta d\Bar{\delta} \;.
\end{equation}
\\
Then, for an arbitrary Borel set $X = I \cross N \subset \mathbb{R}^4 \cross SL(2, \mathbb{C})$, and a bounded, positive observable $A$:

\begin{eqnarray}
     B(X) := \int_X \alpha_{\{a, S\}}(A) dad\mu(S) = \nonumber \\
     \int_I \int_N \alpha_{\{a, S\}}(A) \frac{1}{|\delta|^2}d\beta d\overline{\beta} d\gamma d\overline{\gamma} d\delta d\overline{\delta}da \;.
\end{eqnarray}
\\
Now, the same reasoning done in the Poincaré case applies here, and we have:

\begin{eqnarray}
    P(X) := B^{-\frac{1}{2}} \left[\int_X \alpha_{\{a, S\}}(A) d\nu(\{a, S\})\right] B^{-\frac{1}{2}} = \nonumber \\
    = B^{-\frac{1}{2}}\left[\int_N U(0, S)^*\left(\int_I U(a, \mathds{1})^*AU(a, \mathds{1}) da\right) \times \right. \nonumber \\ 
    \left. U(0, S) \frac{1}{|\delta|^2}d\beta d\overline{\beta} d\gamma d\overline{\gamma} d\delta d\overline{\delta}\right]B^{-\frac{1}{2}} \;.\nonumber
\end{eqnarray}
\\
with the associated time operator given by:

\begin{equation}
    T_A = \int_{\mathbb{R}^4}\int_{SL(2, \mathbb{C})} t P(d\{a, S\}) \;,
\end{equation}
\\
where $d\{a, S\} = \frac{1}{|\delta|^2}d\beta d\overline{\beta} d\gamma d\overline{\gamma} d\delta d\overline{\delta}da$.
\\
This was done, in particular, inspired by the question of Newton-Wigner localization operators \cite{Newton}. The issue that sparked this discussion was trying to create what could be understood as a Newton-Wigner spacetime localization operator. In order to do this, the idea was to create a time operator which was covariant with respect to causal Poincaré transformations, and from it, begin to tie it up with the previous, already known spatial localization operators. After this work, one particular line of efforts that could be pursued, that would directly lead to such spacetime localization operators, would be to extend this method firstly to non-positive observables, and afterwards, to try and understand how such a method could be transplanted in adequate manner to unbounded observables. With these cases being established, a temporal Newton-Wigner localization operator could be constructed directly using the spatial localization operators as basis, using the method studied in this work. The spacetime localization operator would thus be of the form $X^{\mu}$, with $X^0$ the previous time localization operator, and $X^i$ the usual Newton-Wigner operators.
\\

\section{Conclusion}

This work was an in-depth study and a slight extension of a method proposed by Brunneti and Fredenhagen in \cite{Brunetti2002.1}. This was done in order to better understand how the question of time in quantum mechanics can be addressed, and was focused in trying to understand how to approach this problem in more complex quantum systems, such as relativistic ones. This was born out of an attempt into using POVM formalism so as to extend the current spatial Newton-Wigner operators to a spacetime localization operator, which would allow for exploration of the question of localization of systems in time. With the process for constructing covariant POVMs with respect to causal Poincaré transformations done for positive, bounded observables, the next steps towards the previously mentioned goal of building spacetime localization operators would then be to try and study possible ways of either extending upon, or altering \cite{Brunetti2002.1} methods in order to include both not necessarily positive operators, and also unbounded operators. In case these are possible, a spacetime Newton-Wigner operator would come directly from applying this method to the current spatial Newton-Wigner operators.

\begin{acknowledgments}
This work has also been supported by the Conselho Nacional de Desenvolvimento Científico e Tecnológico - Brasil (CNPq), under the protocol 133890/2020-1.
\end{acknowledgments}

%\nocite{*}

\bibliography{references}% Produces the bibliography via BibTeX.

\end{document}